\documentclass[pra,superscriptaddress,showpacs,amsmath,twocolumn]{revtex4}  
\newcommand{\beq}{\begin{equation}}
\newcommand{\enq}{\end{equation}}

\usepackage{graphicx}

\begin{document}

\title{Beyond leading order logarithmic scaling in the catastrophic
self-focusing (collapse) of a laser beam in Kerr media
}

\author{Pavel M. Lushnikov, Sergey A. Dyachenko and Natalia Vladimirova}
\affiliation{Department of Mathematics and Statistics, University of New Mexico, Albuquerque, NM 87131, USA}
\email{plushnik@math.unm.edu}

\date{
\today
}

\begin{abstract}
We study the catastrophic stationary self-focusing (collapse) of laser
beam in nonlinear Kerr media.  The width of a self-similar solutions
near collapse distance $z=z_c$ obeys $(z_c-z)^{1/2}$ scaling law with
the well-known leading order modification of loglog type $\propto
(\ln|\ln(z_c-z)|)^{-1/2}$.  We show that the validity of the loglog
modification requires double-exponentially large amplitudes of the
solution $\sim {10^{10}}^{100}$, which is unrealistic to achieve in
either physical experiments or numerical simulations. We derive a new
equation for the adiabatically slow parameter which determines the
system self-focusing across a large range of solution amplitudes.
Based on this equation we develop a perturbation theory for scaling
modifications beyond the leading loglog.  We show that for the initial
pulse with the optical power moderately above ($\lesssim 1.2$) the
critical power of self-focusing, the new scaling agrees with numerical
simulations beginning with amplitudes around only three times above of
the initial pulse.
\end{abstract}

\pacs{42.65.Jx, 42.65.-k, 52.38.Hb}

\maketitle

\section{Introduction and the main result}
\label{sec:introduction}

The catastrophic collapse (self-focusing) of a high power laser beam has
been routinely observed in nonlinear Kerr media since the advent of
lasers~\cite{AskaryanSovPhysJETP1962,ChiaoGarmireTownesPRL1964,BoydNonlinearOpticsBook2008,SulemSulem1999}.
The propagation of a laser beam
through the Kerr media is described by the nonlinear
Schr\"odinger  equation (NLSE) in dimensionless form,
\begin{eqnarray}\label{nlsl}
  i\partial_z\psi+\nabla^2\psi+|\psi|^{2}\psi=
  0,
\end{eqnarray}
where the beam is directed along $z$-axis, ${\bf r}\equiv (x,y)$ are the
transverse coordinates, $\psi({\bf r},z)$ is the envelope of the
electric field, and $\nabla\equiv\left ( \frac{\partial}{\partial x},
\frac{\partial}{\partial y}\right )$.
NLSE \eqref{nlsl} also describes the dynamics of attractive
Bose-Einstein condensate (BEC) \cite{PitaevskiiStringariBook2003} ($z$
is replaced by the time variable in that case).  In addition, NLSE
emerges in numerous optical, hydrodynamic, and plasma problems, and
describes the propagation of nonlinear waves in general nonlinear
systems with cubic nonlinearity.

If only one transverse coordinate is taken into account, then NLSE
is integrable by the inverse scattering transform
\cite{ZakharovShabatJETP1972} leading to global existence for all solutions (solutions exists for all $z$).
A solution of NLSE which depends on both transverse
coordinates $(x,y)$ can develop a singularity (``blow up'') such that the
amplitude of the solution reaches infinity in a finite distance $z_c$.
Since the blow up is accompanied by dramatic contraction of the
spatial extent of function $\psi$, it is called ``wave collapse'' or
simply ``collapse''~\cite{VlasovPetrishchevTalanovRdiofiz1971,ZakharovJETP1972}.
Near the singularity $z=z_c$, NLSE looses applicability, and
either dissipative or non-dissipative effects must be taken into
account.  Such effects can include the optical damage and formation of
plasma in the Kerr media, inelastic scattering in the BEC, or plasma density
depletion in high temperature laser-plasma interactions
\cite{LushnikovRosePRL2004,LushnikovRosePlasmPysContrFusion2006}.

Equation \eqref{nlsl} can be rewritten in the Hamiltonian form
\begin{eqnarray}
i\psi_t=\frac{\delta H}{\delta \psi^*}
\end{eqnarray}
with the Hamiltonian
\begin{equation}\label{nls1}
 H=\int \big ( |\nabla
 \psi|^2-\frac{1}{2}|\psi|^{4}\big )d {\bf r}.
\end{equation}Another conserved quantity, $N\equiv \int |\psi|^{2}d {\bf r}$, has
the meaning of the optical power (or the number of particles in the BEC).
The sufficient condition for the collapse is $H<0$, while the
necessary condition is $N>N_c$, where $N_c$ is the critical power
 defined below.

While the large power $N\gg N_c$ typically produces multiple collapses
(multiple filamentation of the laser beam
\cite{BergeEtAlMultipleFilamentationPRL2004}) with strong turbulence
behavior~\cite{DyachenkoNewellPushkarevZakharovPhysicaD1992,LushnikovVladimirovaOptLett2010},
the dynamics of each collapsing filament is universal and can be
considered independently.  Each collapsing filament carries the power
$N$ only moderately above $N_c$. We consider a single collapsing
filament (laser beam) centered at ${\bf r}=0$.  For $z\to z_c$ the collapsing
solution of NLSE quickly approaches the cylindrically
symmetric solution, which is convenient to represent through the
following change of variables \cite{SulemSulem1999}:
\begin{eqnarray}\label{selfsimilar}
\psi(r,z)=\frac{1}{L}V(\rho,\tau)e^{i\tau +iL L_z
\rho^2/4}, \quad |{\bf r}|\equiv r,
\end{eqnarray}
Here, $L(z)$ is the $z$-dependent beam width, and
\begin{eqnarray}\label{blowupvar}
\rho=\frac{r}{L}, \quad \tau=\int^z_0 \frac{dz'}{L^2(z')}
\end{eqnarray}
are blow up variables such that $\tau\to \infty$ as $z\to z_c$. Transformation~\eqref{selfsimilar}
was inspired by the discovery of the additional conformal symmetry of
NLSE which is called the ``lens transform''
\cite{TalanovJETPLett1970,KuznetsovTuritsynPhysLettA1985,FibichPapanicolaouSIAMJApplMath1999}.

It follows from  \eqref{nlsl}, \eqref{selfsimilar} and \eqref{blowupvar} that $V(\rho,\tau)$ satisfies
\begin{eqnarray}\label{Vfulleq1}
i\partial_\tau V+ \nabla_\rho^2V-V+|V|^{2}V +\frac{\beta}{4}\rho^2 V=0, \,
\end{eqnarray}
where
\begin{eqnarray}\label{betadef}
\beta=-L^3 L_{zz} \quad \mbox{and} \quad \nabla_\rho^2\equiv \partial^2_\rho+{\rho}^{-1}\partial_\rho.
\end{eqnarray}
As $z\to z_c$, $\beta$ approaches zero
adiabatically slowly and
$V(\rho)$ approaches the ground state soliton
$R(\rho)$~\cite{FibichPapanicolaouSIAMJApplMath1999}.
The ground state soliton is the radially symmetric, $z$-independent
solution of NLSE, $-R+\nabla ^2_\rho R+R^3=0$.
It is positive definite, i.e., $R>0$, with asymptotic
$R(\rho)=e^{-\rho}[A_R\rho^{-1/2}+O(\rho^{-3/2})],$ $\rho \to \infty,$
$A_R\equiv 3.518062 \ldots$
\cite{FibichPapanicolaouSIAMJApplMath1999}. Also $R$ defines the
critical power
\begin{equation}\label{Ncdef}
N_c\equiv 2\pi\int R^2\rho d\rho=11.7008965\ldots
\end{equation}
The limiting behavior in $V\to R$ as $z\to z_c$ implies that
the $\partial_\tau V$ term in \eqref{Vfulleq1} is a small correction
compare to the other terms. Also $\beta$ can be interpreted as
 quantity
proportional to the excess of particles
above critical, $N-N_c$, in the collapsing region \cite{MalkinPhysD1993,FibichPapanicolaouSIAMJApplMath1999}.

Refs. \cite{FraimanJETP1985} and
\cite{LeMesurierPapanicolaouSulemSulemPhysicaD1988}
 found that the leading order dependence of $L(z)$ has
the following square-root-loglog form
\begin{eqnarray}
L\simeq \left(2\pi \frac{z_c-z}{\ln|\ln(z_c-z)|}\right)^{1/2}.
\label{double}
\end{eqnarray}
(Ref. \cite{FraimanJETP1985} has a "slip of pen" in a final expression, see
e.g. \cite{MalkinPLA1990} for a discussion.)  The validity of the
scaling~\eqref{double} at $z\to z_c$ was rigorously proven in
Ref.~\cite{merleraphael2006}.  However, numerous attempts to verify
the modification of $L\propto ({z_c-z})^{1/2}$ scaling have failed to
give convincing evidence of the loglog dependence (see
e.g.~\cite{KosmatovShvetsZakharovPhysicaD1991,AkrivisDougalisKarakashianMcKinneySIAMJSciComp2003}).
Lack of validity of loglog law was also discussed in Ref. \cite{FibichPapanicolaouSIAMJApplMath1999}.
Note that without logarithmic modification, the scaling $({z_c-z})^{1/2}$ implies
$\beta=const$, $N=\infty$, and infinitely fast rotation of the phase
for $r\to \infty$ with $\beta \neq 0$.  Thus, the logarithmic
modification is necessary and is responsible for  the adiabatically
slow approach of $\beta$ to 0.

A qualitatively similar problem of logarithmic modification of
square-root scaling also occurs in the Keller-Segel equation, which
describes either the collapse of self-gravitating Brownian particles
or the chemotactic aggregation of micro-organisms
\cite{HeVe1996a,VelazquezSIAMJApplMath2002,LushnikovPhysLettA2010,DejakLushnikovOvchinnikovSigalPhysD2012,DyachenkoLushnikovVladimirovaAIP2011}.
 It was shown in~\cite{DyachenkoLushnikovVladimirovaAIP2011,DyachenkoLushnikovVladimirovaKellerSegelArxiv2013} that the
leading logarithmic modification in Keller-Segel equation is valid
only for very large amplitudes ($\gtrsim 10^{10000}$).
Also in~\cite{DyachenkoLushnikovVladimirovaAIP2011,DyachenkoLushnikovVladimirovaKellerSegelArxiv2013}, the
perturbation theory was developed beyond the leading
order logarithmic correction. That theory was shown to be accurate starting from
moderate amplitudes ($\gtrsim 3$) of collapsing solution.

Following qualitatively some ideas of~\cite{DyachenkoLushnikovVladimirovaAIP2011,DyachenkoLushnikovVladimirovaKellerSegelArxiv2013},
in this paper we develop the
perturbation theory about the self-similar solution of
\eqref{Vfulleq1} with $V\simeq R(\rho)$ and show that the scaling
\eqref{double} dominates only for very large amplitudes
\begin{eqnarray}
\label{doubleexponent}
|\psi|\gtrsim {10^{10}}^{100}.
\end{eqnarray}
Instead of pursuing this unrealistic limit, we
suggest a following new expression (derived below) as a practical choice for
the experimental and theoretical study of self-focusing:
\begin{widetext}
\begin{align}
\label{Ltot}
\begin{split}
&L=[{2\pi(z_c-z)}]^{1/2}
\left (\ln{A}-4\ln 3+4\ln{\ln{A}}
+\frac{4 (-1-4 \ln{3}+4\ln{\ln{A}})}{\ln{A}} \right.  \\
&\left .  +\frac{-28-80 \ln{3}-32 (\ln{3})^2-\pi ^2 c_1
+80 \ln{\ln{A}}+64 (\ln{3}) \ln{\ln{A}}-32 [\ln{\ln{A}}]^2}{(\ln{A})^2}  \right )^{-1/2}, \\
&A=-3^4\frac{\tilde M}{2\pi^3}\ln{\left[ \left [2\pi(z_c-z)\right ]^{1/2}\frac{e^{-b_0}}{L(z_0)}\right ]}, \quad
    \ \tilde M=44.773 \ldots,\quad \beta_0=\beta(z_0),
\ c_1=4.793\ldots, c_2=52.37\ldots,\\
&b_0=\frac{e^{\frac{\pi }{\sqrt{\beta_0 }}}}{\tilde M} \left(\frac{2 \beta_0 ^2}{\pi }+
\frac{8 \beta_0 ^{5/2}}{\pi ^2}+\frac{2 \beta_0 ^3 \left(20
+\pi ^2 c_1\right)}{\pi ^3}+\frac{12 \beta_0 ^{7/2} \left(20 \pi ^3+\pi ^5 c_1\right)}
{\pi ^7}+\frac{2 \beta_0 ^4 \left(840 \pi ^3+42 \pi ^5 c_1+\pi ^7 c_2\right)}{\pi ^8}\right).
\end{split}
\end{align}
\end{widetext}
This expression depends on an additional parameter, $z_0,$ defined  below.
$L(z)$ is only weakly sensitive to the choice of  $z_0<z_c$ provided $z_0$ is  larger  than
 the smallest distance
at which the collapsing solution has approximately reached the self-similar form.
\begin{figure}
\begin{center}
\includegraphics[width =2.85 in]{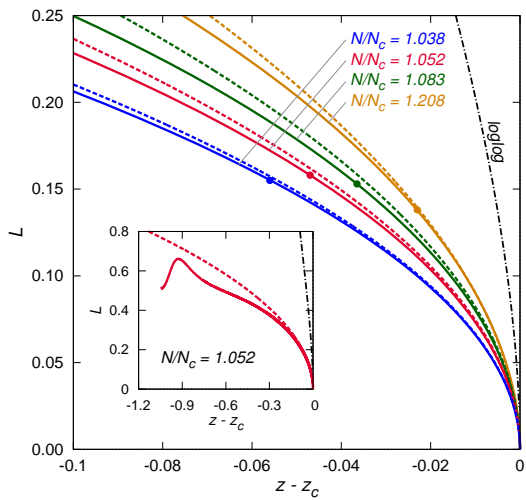}
\end{center}
\caption{(Color online) A dependence of the beam width $L$ on $z-z_c$
  obtained from numerical simulations of NLSE~\eqref{nlsl} (solid
  lines) and from equation~\eqref{Ltot} (dashed lines) for different
  initial conditions.  Each pair of closely spaced solid and dashed
  lines corresponds to the same Gaussian initial condition
  $\psi(r,0)=pe^{-r^2}$.  The curves are labeled by the power $N=\pi
  p^2/2$ (scaled by the critical power $N_c$).  The dash-dotted line
  shows $L$ from the loglog law~\eqref{double}.  The dashed lines are
  obtained from equation~\eqref{Ltot} using the parameters
  $L_0=L(z_0)$ and $\beta_0=\beta(z_0)$ taken from numerical
  simulations at locations $z=z_0$. These locations are marked by the
  thick dots at each solid line.  These values of $z_0$ are chosen by
  the criterion $[\max_{\bf r}|\psi({\bf r},z_0)|]/[\max_{\bf
      r}|\psi({\bf r},0)|]=5.$ The inset shows $L(z)$ for
  $N/N_c=1.052$ starting from the beginning of simulation, $z=0$. It
  is seen in the inset
  that about 2-fold decrease of $L$ compare with the initial value $L(0)$ already
  produces a good agreement between the simulation of
  NLSE~\eqref{nlsl} and equation~\eqref{Ltot}.
 }
\label{fig:Lt}
\end{figure}

\begin{figure}
\begin{center}
\includegraphics[width = 2.85 in]{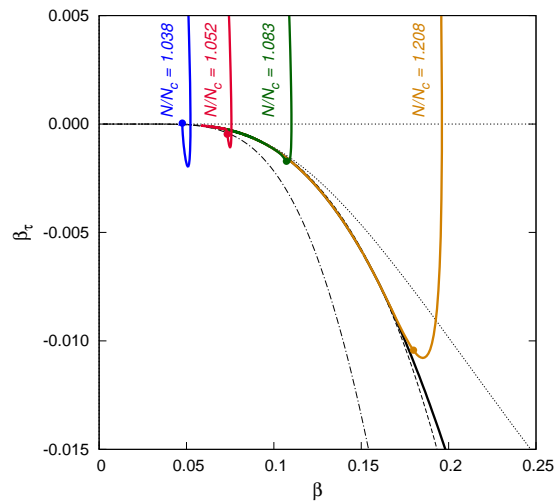}
\end{center}
\caption{(Color online) 
  lines show $\beta_\tau(\beta)$ from numerical simulations of NLSE
  \eqref{nlsl} with the same initial conditions as in
  Figure~\ref{fig:Lt}.  The curves are labeled by the values of
  $N/N_c$. It is seen that the solid curves converge to a single
  universal $\beta_\tau(\beta)$ curve after the initial transient
  evolution.
   The universal curve is independent on initial conditions.
  Similar to Figure~\ref{fig:Lt},
  the thick dots mark the locations of $z=z_0$ at each solid line,
  i.e.  they indicate the pairs of points
  $(\beta(z_0),\beta_\tau(z_0))$.  The dashed line corresponds to
  $\beta_\tau(\beta)$ from \eqref{betatauNb1}. 
  obtained either numerically.  See also the text for the description
  of the dash-dotted and dotted lines.
}
\label{fig:betataubeta}
\end{figure}

\begin{figure}
\begin{center}
\includegraphics[width = 2.85 in]{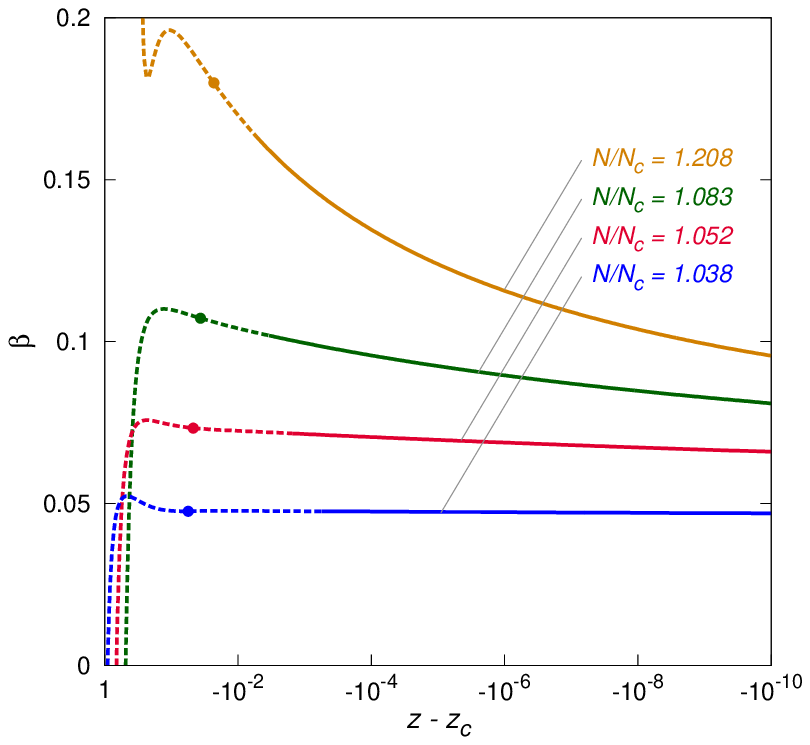}
\end{center}
\caption{(Color online)  Dependence of $\beta$ on $z-z_c$ for the same set of
  simulations as in Figure~\ref{fig:Lt}.   The initial fast
  evolution is responsible for the formation of the quadratic phase
  (see equation~\eqref{selfsimilar}) and is specific to our Gaussian initial conditions with zero phase.  The evolution slows
  down after $\beta$ passes through the local maximum; the following
  change in $\beta$ is especially slow for smaller values of $N/N_c$.
  The transitions from dashed to solid lines indicate the collapse of
  the corresponding $\beta_\tau(\beta)$ curves  onto the single
  universal curve shown in Figure~\ref{fig:betataubeta}.  The relative
  difference of $10^{-3}$ between a particular simulation curve and the
  universal curve is used as a transition criterium.  Similar to
  Figure~\ref{fig:Lt}, the thick dots mark the locations of $z=z_0$.
}
\label{fig:betazzc}
\end{figure}

To illustrate the poor agreement with the loglog law at moderate
amplitudes, Figure~\ref{fig:Lt} shows the dynamics of $L(z)$ obtained
from numerical simulations.  The simulations were started with
different initial conditions in the form of Gaussian beams $\psi({\bf
  r},0)=pe^{-r^2}$ with the power $N=\pi p^2/2$.  Figure \ref{fig:Lt}
shows that $L(z)$ neither agrees with the loglog law~\eqref{double}
nor it is universal.  In contrast, the dependence of $\beta_\tau$ on
$\beta$ appears to be universal as demonstrated in
Figure~\ref{fig:betataubeta}.  The curves corresponding to different
initial conditions converge to a single $\beta_\tau(\beta)$ curve
after the initial transient evolution. The resulting single curve is
universal and independent on initial conditions.  This universality is
the key for the analytical theory developed below.  Note that the
dependence of $\beta$ on $ z-z_c$ is also not universal as seen in
Figure~\ref{fig:betazzc} so it cannot be used effectively for the
development of the analytical theory.

\begin{figure}
\begin{center}
\includegraphics[width =2.85 in]{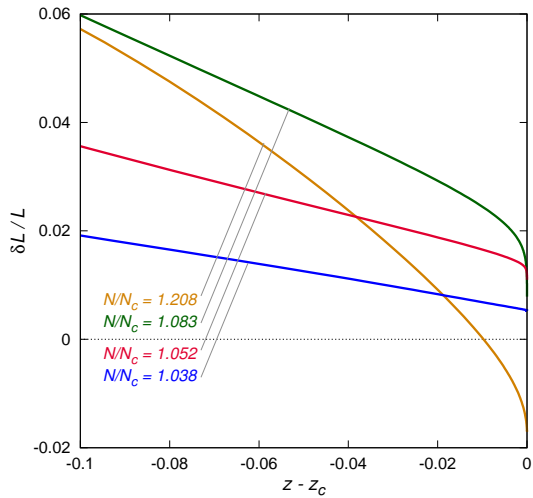}
\end{center}
\caption{(Color online) The relative error, $\delta L/L$, between
  $L(z)$ obtained from the numerical simulations of NLSE~\eqref{nlsl}
  and $L(z)$ from equation~\eqref{Ltot} for the same set of
  simulations as in Figure \ref{fig:Lt}.  It is seen that the relative
  errors decreases as $(N-N_c)/N_c$ approaches zero. The exception is
  the curve for much larger value $N/N_c=1.208$, where
  equation~\eqref{Ltot} is formally on the boundary of its range of
  applicability.  }
\label{fig:Ltrelat}
\end{figure}

Figure~\ref{fig:Lt} also demonstrates the excellent agreement between
the analytical expression~\eqref{Ltot} and numerical simulations of
NLSE~\eqref{nlsl}.  Figure ~\ref{fig:Ltrelat} shows the relative error
between $L(z)$ obtained from the numerical simulations of
NLSE~\eqref{nlsl} and $L(z)$ from equation~\eqref{Ltot}.
The relative errors decreases with the decrease of
$(N-N_c)/N_c$.
The only exception is the curve for a significantly larger power $N/N_c=1.208$
which is formally beyond the applicability of equation~\eqref{Ltot}.
Equation~\eqref{Ltot} is derived in the limit $(N-N_c)/N_c\to 0$, as explained below.
However, even in the case of $N/N_c=1.208$  the relative error $\lesssim 6\%$.
 In evaluating ~\eqref{Ltot} we
used the parameters $L_0=L(z_0)$ and $\beta_0=\beta(z_0)$ taken from
numerical simulations at locations $z=z_0$.
These locations are shown by  the thick dots in Figure~\ref{fig:Lt}.
 Similar, the thick dots show the
corresponding points $\beta(z_0)$ and $\beta_\tau(z_0)$ in Figures
\ref{fig:betataubeta} and \ref{fig:betazzc}.
We choose $z_0$ as the
propagation distance where the amplitude of collapse exceeds the
initial amplitude of Gaussian pulse by a factor of five, i.e.
$[\max_{\bf r}|\psi({\bf r},z_0)|]/[\max_{\bf r}|\psi({\bf r},0)|]=5.$
Choosing $z_0$ larger
than defined above (e.g. by 10 fold increase of collapse amplitude)
results only in very small variations ($\lesssim 0.2\%$) of dashed
lines in Figure~\ref{fig:Lt} for $N/N_c\lesssim 1.1$. It means that
the prediction of analytical expression is only weakly dependent on
$z_0$.

The paper is organized as follows.  In
Section~\ref{sec:amplitudeequations} we  approximate
the collapsing solution by the expansion about the soliton solution in
blow-up variables.  The perturbations of  this solution
determines the rate of collapse which allows us to derive the reduced
ordinary differential equation (ODE) system for unknowns $L(z)$ and
$\beta(z)$. In Section \ref{sec:blowuprate} we find the asymptotic
solution of  this reduced system in the limit $z\to
z_c$ and derive the scaling~\eqref{Ltot}.  In Section
\ref{sec:experiementnumerics} we estimate the range of applicability
of a NLSE collapsing solution in experiment.  In Section
\ref{section:simulationalgorithm} we briefly describe
  the method used in the NLSE simulation, as well as we discuss the
procedure for the extraction of the parameters of the collapsing solutions
$\beta(z)$ and $L(z)$ from the simulations.  In Section
\ref{sec:conclusion} the main results of the paper are discussed.


\section{Reduction of NLSE collapsing solution to ODE  system for  $L(z)$ and $\beta(z)$}
\label{sec:amplitudeequations}

To determine $\beta_\tau(\beta)$ analytically, we consider the ground
state soliton solution $V_0(\beta,\rho)$ of \eqref{Vfulleq1} given by
\begin{eqnarray}\label{V0eq1}
 \nabla_\rho^2V_0-V_0+V_0^3 +\frac{\beta}{4}\rho^2 V_0=0.
\end{eqnarray}
The function $V_0(\beta,\rho)$ has an oscillating tail, $V_0(\beta,\rho)=c
{\rho^{-1}}\cos{\left [\frac{\beta^{1/2}}{4}\rho^2 -\beta^{-1/2}\ln
\rho+\phi_0\right ]}+O(\rho^{-3}),$ with $c, \phi_0=const$ and $\rho \gg
2/\beta^{1/2}$.  Here, by ground state soliton $V_0$, we mean the real
function 
such that it minimizes $|c|$ in the tail. It implies that $V_0$ has
only small amplitude oscillations with $|c|\ll 1$ for $0<\beta\ll 1.$

The full solution $V(\beta,\rho)$ of \eqref{Vfulleq1} is well approximated
by $V_0(\beta,\rho)$ for $\rho \lesssim 1$. However, the small but
nonzero value of $\partial_\tau V_0=\beta_\tau \frac{d V_0}{d\beta}$
provides an imaginary contribution to $V$.  To account for the imaginary
contribution at the leading order, we allow $V_0$ to be complex
(replacing it by $\tilde V_0$), similar to the approach
of~\cite{DyachenkoNewellPushkarevZakharovPhysicaD1992,FibichPapanicolaouSIAMJApplMath1999}.
We formally add an exponentially small term $i\nu(\beta) \tilde V_0$
to~\eqref{V0eq1} as follows, $\nabla^2\tilde V_0-\tilde V_0+|\tilde
V_0|^{2}\tilde V_0 +\frac{\beta}{4}\rho^2 \tilde V_0-i
\nu(\beta) \tilde V_0=0$.
The yet unknown $\nu(\beta)$ accounts for the loss of power of $
\tilde V_0$ by emission into the tail. One can reinterpret the resulting
equation as a linear Schr\"odinger equation with a self-consistent
potential $U\equiv-|\tilde V_0|^2-\frac{\beta}{4}\rho^2$ and a complex
eigenvalue $E\equiv-1-i\nu(\beta)$. (This type of nonself-adjoint
boundary value problems was introduced by Gamov in 1928 in the theory
of $\alpha$-decay \cite{LandauLifshitzQuantMech1981}.)  Assuming
$\beta\ll 1$, we identify two turning points, $\rho_a\sim
1$ and $\rho_b\simeq 2/\beta^{1/2}$, at which $Re(E)+U=0$.
Using the WKB  (Wentzel-Kramers-Brillouin) approximation we consider the
tunneling  from the collapsing region $\rho\lesssim 1$
 through the classically forbidden region
$\rho_a<\rho<\rho_b$, and obtain, similar to
\cite{DyachenkoNewellPushkarevZakharovPhysicaD1992} that
\begin{eqnarray}\label{Vasymp}
\tilde V_0= e^{-\frac{\pi}{2\beta^{1/2}}}
\exp{\left [i\frac{\beta^{1/2}}{4}\rho^2 -i\beta^{-1/2}\ln \rho-i\tilde \phi_0\right ]} \nonumber \\
\times \frac{2^{1/2}A_R}{\beta^{1/4}} [\rho^{-1}+O(\rho^{-3})],
\quad \tilde\phi_0=const, \quad \rho \gg \rho_b,
\end{eqnarray}
where $A_R$ results from the matching of the asymptotic of $R$ with
the WKB solution. We also note that the tail
\eqref{Vasymp} is derived in the adiabatic approximation which is
valid for large but finite values of radius, $2/\beta^{1/2}\ll r/L \ll
A\,( 2/\beta^{1/2})$, where $A(z)\gg 1$ is a slowly changing factor in
comparison with $L(z)$.  Even though for $r/L \gtrsim A\,(
2/\beta^{1/2})$ the solution is not self-similar
\cite{MalkinPLA1990,BergePesmePLA1992,BergePhysRep1998}, its
large-radius asymptotic has no influence of $L(z)$ and is not
considered here.

We define the power (the number of particles) $N_b$ in the collapsing
region $\rho<\rho_b$ as %
\begin{eqnarray}\label{Nbdef}
N_b=\int \limits_{r<\rho_bL}|\psi|^2 d{\bf r}=2\pi \int \limits_{\rho<\rho_b}|V|^2 \rho d\rho.
\end{eqnarray}
and a flux $P$ beyond the second turning point
$\rho_b=2/\sqrt{\beta}$ as $P=2\pi\rho \left [i VV^*_\rho +c.c. \right
]|_{\rho=\rho_b},$ where c.c. stands for complex conjugate terms.
From conservation of $N$, the flux $P$ determines the change of $N_b$ as
\begin{eqnarray}\label{betatauNb}
\frac{d N_b}{d\tau}=
 -2\pi\rho \left [i VV^*_\rho +c.c. \right ], \ \rho\gg \rho_b,
\end{eqnarray}
where we approximated $P$ at $\rho=\rho_b$ through its value at
$\rho\gg \rho_b$ taking advantage of almost constant flux to the right of the
second turning point. Using the adiabatic assumption that
%
$
\frac{d N_b}{d\tau}=\beta_\tau \frac{d N_b}{d\beta},
$
%
%
and approximating $V$ in \eqref{betatauNb} by \eqref{Vasymp} we obtain that
\begin{eqnarray}\label{betatauNb1}
\beta_\tau=-4\pi A_R^2\left (\frac{d N_b}{d\beta}\right )^{-1}e^{-\frac{\pi}{\beta^{1/2}}} .
\end{eqnarray}
Recalling the definition of $\nu(\beta)$, one can also find
$\nu(\beta)\simeq (2\pi
A_R^2/N_b)e^{-\frac{\pi}{\beta^{1/2}}}$ from~\eqref{betatauNb1}.

\begin{figure}
\begin{center}
\includegraphics[width = 2.65 in]{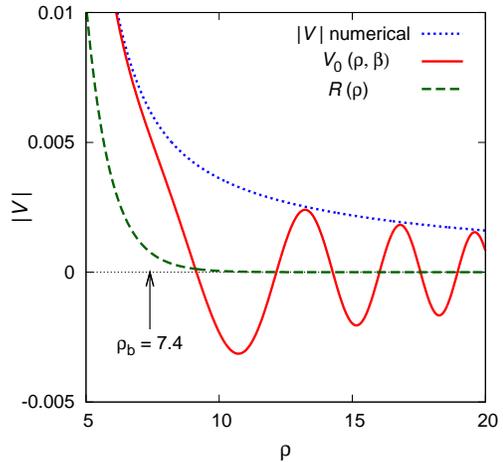}
\end{center}
\caption{(Color online)
Asymptotic $\rho \gg 1$ for $V_0$ (solid line), full numerical
solution $|V|$ (dotted line) and $R$ (dashed line) for $\beta
=0.073$. It is seen that $V_0$ and $|V|$ almost coincide for
$\rho<\rho_b.$ }
\label{fig:psiasymp}
\end{figure}

The next step is to find $\frac{d N_b}{d\beta}$ in \eqref{betatauNb1}.
We based our derivation on a crucial observation that the absolute
value $|V(\beta,\rho)|$ of the numerical solution of \eqref{Vfulleq1}
coincides with $V_0(\beta,\rho)$ for $0\le \rho\lesssim\rho_b$, as
shown in Figure~\ref{fig:psiasymp}.  
the approximation $V_0(\beta,\rho)\simeq R(\rho)+dV(\beta,\rho)/d
\beta|_{\beta=0}$ used previously (see
e.g.~\cite{FibichPapanicolaouSIAMJApplMath1999}) is limited to $\rho
\ll \rho_b$ because the amplitude $c$ of the tail of $V_0$ has the
essential complex singularity $c\propto e^{-\pi/(2\beta^{1/2})}$ for
$\beta\to 0.$ Approximating $N_b$ through replacing $V$ in
\eqref{Nbdef} by $V_0(\beta,\rho)$ we obtain the following series
\begin{eqnarray}\label{betaNbdefser}
\frac{d N_b}{d\beta}\! =\! 2\pi M\! \left [1+c_1\beta+c_2\beta^2+c_3\beta^3+c_4\beta^4+c_5\beta^5 \right ],
\end{eqnarray}
where $M\equiv (2\pi)^{-1}dN_b/d\beta|_{\beta=0} =
(1/4)\int^\infty_0\rho^3R^2(\rho)d\rho=0.55285897\ldots$ and
coefficients $c_1=4.74280, \, c_2=52.3697,\, c_3=297.436, \,
c_4=-4668.01, \, c_5=10566.2$ are estimated from the numerical
solution of~\eqref{V0eq1}.  Here the value of $c_1$ is obtained from
the numerical differentiation: $c_1=(2\pi
M)^{-1}d^2N_b/d\beta^2|_{\beta=0}$.  One can in principle find
coefficients $c_2,c_3,\ldots$ from higher order numerical differentiation
at $\beta=0$. However, the radius of convergence of the corresponding
Taylor series is $\beta\sim 0.04$.  Yet the range of $\beta$ resolved
in our NLSE simulations is $\beta\gtrsim 0.05$ as seen in
Figure~\ref{fig:betataubeta}.  Thus it would be inefficient to use the
Taylor series (centered at $\beta=0$) to approximate $\frac{d
  N_b}{d\beta}$ in \eqref{betaNbdefser} for $\beta\gtrsim 0.05$.
Instead we approximate $c_2,\ldots,c_5$ from the polynomial fit in the
range $0.0< \beta< 0.23$.
This procedure gives the numerical values
given above.  The relative error between the exact value
of $\frac{d N_b}{d\beta}$ and the polynomial interpolation
\eqref{betaNbdefser} is $<1.6\%$ in the range $0\le \beta <0.23$.
If only $c_1$ and $c_2$ are taken into account in
\eqref{betaNbdefser}, then the relative error is $<1.0\%$ in the range
$0\le \beta <0.09$.  Figure~\ref{fig:betataubeta} shows that
equations~\eqref{betatauNb1} and~\eqref{betaNbdefser} approximate well
the full numerical solution for $\beta\lesssim 0.18$.  Indeed,
$\beta_\tau(\beta)$ from \eqref{betatauNb1} with $dN_b/d\beta$,
obtained either numerically via $V_0(\beta,\rho)$ or by using
equation~\eqref{betaNbdefser}, are indistinguishable on the plot (they
are both shown by the dashed line).  The dotted line corresponds to
equation~\eqref{betaNbdefser} with only $c_1$ and $c_2$ taken into
account.

For comparison, the dash-dotted line in Figure~\ref{fig:betataubeta}
shows
the standard approximation for $\beta_\tau(\beta)$
\cite{FibichPapanicolaouSIAMJApplMath1999}, which corresponds
to~\eqref{betaNbdefser} with the expression in square
brackets replaced by~1.
As we see, the standard approximation fails all way
down to $\beta \approx 0.05$. Further decrease of $\beta$
is unresolvable in our simulations (which typically
reach $\max |\psi| \sim 10^{15}$).

From equations~\eqref{blowupvar},\eqref{betadef},\eqref{betatauNb1} and
\eqref{betaNbdefser} we obtain a closed system
\begin{subequations}\label{betaLsyst}
\begin{align}
 \frac{d\beta}{d\tau}&= - \frac{2A_R^2}{M(1+c_1\beta+c_2\beta^2+c_3\beta^3+c_4\beta^4+c_5\beta^5) } \nonumber \\
&\qquad\qquad\qquad\qquad\qquad\qquad\times e^{-\frac{\pi}{\beta^{1/2}}} , \label{betaLsysta}\\
\frac{d^2 L}{dz^2}&=-L^{-3}\beta,\label{betaLsystb} \\
\frac{d \tau}{dz}&=L^{-2}, \label{betaLsystc}
\end{align}
\end{subequations}
from which the unknowns $\beta(z)$ and $L(z)$ can be determined.  This
system is the ODE system for independent variable $z$ because $\tau$
can be easily excluded from the system using equation
\eqref{betaLsystc}.


\section{Asymptotic solution of the reduced system}
\label{sec:blowuprate}

In this section we look for the asymptotic solution of the reduced
system \eqref{betaLsyst}, in the limit $z\to z_c, \ \tau\to \infty$,
$\beta\to 0$ and $L\to 0$ to derive our main result,
equation~\eqref{Ltot}.  We introduce the adiabatically slow variable
\begin{equation}\label{adef1}
a = -L\frac{d L}{d z},
\end{equation}
which is also expressed through $\tau$ as $a=-L^{-1}\frac{dL}{d \tau}$
according to \eqref{betaLsystc}.  Here and below, we use the same
notations for all functions with the same physical meaning,
independently of their arguments: $L=L(z)=L(\tau)=L(\beta)$, etc.

Using equations~\eqref{betaLsystc} and \eqref{adef1} we obtain that
\begin{equation}\label{betaarel}
\beta =a^2+a_\tau.
\end{equation}
However, the adiabatic slowness of $a$ requires $a_\tau\ll a^2$ because,
by the chain rule of differentiation, $a_\tau=a_\beta \beta_\tau$ while
$ \beta_\tau$ is exponentially small for $\beta\ll 1$, as follows from \eqref{betaLsysta}.
Then at the leading order we obtain from \eqref{betaarel} that
\begin{equation}\label{abetareduced}
a=\beta^{1/2}.
\end{equation}
Using \eqref{adef1} we obtain that $dz=-\beta^{-1/2} LdL$, which
allows us to explicitly integrate \eqref{betaLsysta} in terms of
variables $L$ and $\beta $ and their initial values $L_0=L(z_0)$,
$\beta_0(z_0)$ ($z_0$ is defined above).  The explicit expression is
cumbersome and includes the exponential integral function
$Ei(\pi/\beta^{1/2}),$ $Ei(x)=-\int^\infty_{-x}e^{-y}y^{-1}dy$. We
asymptotically expand this expression for $\pi/\beta^{1/2} \gg 1$ to
obtain the following expression
\begin{equation}\label{lnL0Lbetabeta0}
\begin{split}
 -\ln{\frac{L}{L_0}}=\frac{2\pi^3 e^x}{\tilde M}\left [\frac{1}{x^4}+\frac{4}{x^5}+\frac{20+\pi ^2 c_1}{x^6} \right. \qquad\qquad
  \\
  \left.   +\frac{120+6 \pi ^2 c_1}{x^7}+\frac{840+42 \pi ^2 c_1+\pi ^4 c_2}{x^8}\right.\\
 + \left .\left .O\left(\frac{1}{x^{9}} \right )
\right ]\right |^x_{x=\pi/\beta_0^{1/2}}, \ x\equiv \frac{\pi}{\beta^{1/2}}, \ \tilde M=\frac{2A_R^2}{M}.
\end{split}
\end{equation}

Addition of the correction term $a_\tau$ in \eqref{betaarel} can be
easily done as a small perturbation.
For the range of parameters
considered in our simulations, such correction would result
in the change of all solutions by $<1\%$.  Therefore the
correction is omitted in this paper.
Deriving~\eqref{Ltot}, we used leading order terms with only $c_1$
and $c_2$  taken into account in equation~\eqref{betaNbdefser}.
(The corresponding $\beta_\tau(\beta)$ is shown by a
dotted line in Figure~\ref{fig:betataubeta}.)  Thus $c_1$ and $c_2$
in~\eqref{betaNbdefser} are sufficient to produce very good agreement
with the simulations shown in Figure~\ref{fig:Lt}.

When equation~\eqref{lnL0Lbetabeta0} is interpreted as an implicit
expression for $x$ as a function of $\ln{\frac{L}{L_0}}$,
it becomes a remote relative of the Lambert W-function.
Such implicit expression can be solved for $x$ assuming $x\gg 1$ by
iterations as follows:
\begin{equation} \label{phieqsol}
\begin{split}
x=L_1+4L_2-4\ln{3}    +\frac{4(4L_2-1-4\ln{3})}{L_1}\qquad\\
+\frac{16[-2L_2^2+L_2(5+4\ln{3})-2(\ln{3})^2-5\ln{3]}}{L_1^2} \\
+\frac{-28-\pi^2c_1}{L_1^2}+O\left (\frac{L_2^3}{L_1^3}\right ),
\end{split}
\end{equation}
where
\begin{equation}\label{L1L2def}
L_1= \ln {\left [\frac{3^4\tilde M}{2\pi^3}\left(\ln{\frac{L_0}{L}}+b_{0}\right )\right ]}, \quad L_2= \ln{L_{1}},
\end{equation}
with $b_0$ defined in equation \eqref{Ltot} ($b_0$  is proportional to
the
right-hand side of equation \eqref{lnL0Lbetabeta0} with
$x=\pi/\beta_0^{1/2}$).  The factor $3^4$ in the definition of
$L_1$ is somewhat arbitrary: we can multiply both sides of
equation \eqref{lnL0Lbetabeta0} by the arbitrary constant before
starting the interation procedure to derive \eqref{phieqsol}.  This
factor shows up in equation \eqref{phieqsol} through terms with powers
of $\ln 3$. The particular choice of $3^4$ allows us to speed up convergence
of the series expansion \eqref{phieqsol} for not very large values of
$L_1$.

We now introduce the collapse distance $z_c$ into the system \eqref{betaLsyst} as follows
\begin{equation}\label{tctdef}
\begin{split}
 z_c-z=\int \limits^{z_c}_z dz'=-\int\limits^{0}_{L}\frac{L'dL'}{a(L')}=
 \int\limits^{\ln{L}}_{-\infty}\frac{(L')^2d\ln{L'}}{[\beta(L')]^{1/2}},
\end{split}
\end{equation}
where we used \eqref{adef1} and \eqref{abetareduced}.
Using \eqref{phieqsol} we express $\beta$ in   \eqref{tctdef} through $L$.
Then we evaluate the integral in equation \eqref{tctdef} asymptotically for $\ln{L}\to -\infty$ using the Laplace method  (see
e.g. \cite{SidorovFedoryukShabuninComplexAnalysisBook1985,OlverBook1985})
which gives
\begin{widetext}
\begin{eqnarray} \label{tctL1}
&z_c-z=\frac{L^2}{2\pi}\left [L_1-4\ln 3+4L_2
+\frac{4 (-1-4 \ln{3}+4L_2)}{L_1} 
+\frac{-28-80 \ln{3}-32 (\ln{3})^2-\pi ^2 c_1
+80 L_2+64 (\ln{3}) L_2-32 L_2^2}{L_1^2}+O\left(\frac{L_2^3}{L_1^3}\right )  \right ].
\end{eqnarray}
\end{widetext}
We solve \eqref{tctL1} for $L$ by iterations and  obtain
  \eqref{Ltot} at the leading order. In that leading order derivation we neglected the error term
$O(\ldots)$ and used \eqref{L1L2def}.  The asymptotic expansion
\eqref{lnL0Lbetabeta0} is well convergent for $\beta\lesssim 0.1$ only.
It formally limits applicability of \eqref{Ltot}
        to $\beta\lesssim 0.1$.
For the simulation with the largest shown value
$N/N_c=1.208$ we have the condition $\beta\gtrsim 0.1$ as
seen in Figure \ref{fig:betazzc}, i.e. on the border of \eqref{Ltot} applicability at best.
        This explains a relatively poor convergence of the numerical simulation value of $L(z)$ to \eqref{Ltot} for
        $N/N_c=1.208$ as shown in Figure \ref{fig:Ltrelat}.
We note however, that
even in this case the relative error for $L(z)$ is moderately small:
$\lesssim 6\%$. It means that while $N/N_c=1.208$ is formally beyond
the applicability limits of equation ~\eqref{Ltot}, the numerical error
remains moderate and equation ~\eqref{Ltot} can be used
(with caution)  even
beyond its formal applicability condition $\beta\lesssim 0.1$.


\section{Experimental estimates}
\label{sec:experiementnumerics}

In this Section we show that the dynamic range of laser intensities for NLSE
applicability can be made quite large in experiment to allow the experimental verification of the collapse scaling \eqref{Ltot}.
We identify the required ranges of laser intensity, laser pulse duration and laser propagation distance in Kerr media for the robust NLSE applicability in the collapse regime.
We found above that \eqref{Ltot} is applicable after the initial growth of the pulse amplitude by a factor $\sim 2-3$.  It implies that
the laser intensity increases by a factor $\sim 4-9$. For instance, the experimental increase of the laser intensity by $2-3$ orders of magnitude would be more than sufficient to the robust identification/verification
of the collapse scaling  \eqref{Ltot}.
We focus our estimates on the self-focusing of a laser beam in fused silica although our estimates are easy to modify for other Kerr media.
We choose for the estimate that $N/N_c=1.052$ as in the inset of Figure \ref{fig:Lt}. It determines the collapse distance $z_c\simeq 1.047499$ in  dimensionless units.

We first
consider a stationary self-focusing of the laser beam in Kerr medium.
(We assume for now that the pulse duration is long enough to neglect
time-dependent effects. We estimate the range of allowed
pulse durations below.)  NLSE \eqref{nlsl} in dimensional units with added
multi-photon absorbtion (MPA) takes the following form (see e.g.
\cite{BergeSkupinNuterKasparianWolfBergePhysRep2007}):
\begin{eqnarray}\label{nlsdimensionall}
  i\partial_z\tilde\psi+\frac{1}{2k}\nabla^2\tilde\psi+\frac{kn_{2}}{n_0}|\tilde\psi|^{2}
  \tilde\psi+i\frac{\beta^{(K)}}{2}|\tilde\psi|^{2K-2}\tilde \psi=
  0,
\end{eqnarray}
where $k=2\pi n_0/\lambda_0$ is the wavenumber in media, $\lambda_0$
is the vacuum wavelength, $n_0$ is the linear index of refraction, and
$n_2$ is the nonlinear Kerr index.  The index of refraction is
$n=n_0+n_2 I$, where $I=|\tilde \psi|^2$ is the light intensity. Also
$K$ is the number of photons absorbed by the electron in each
elementary process ($K$-photon absorbtion) and $\beta^{(K)}$ is the
multiphoton absorbtion coefficient. For fused silica with
$\lambda_0=790\text{nm}$, $n_0=1.4535$, and $n_2=3.2\cdot
10^{-16}\text{cm}^2/\text{W}$. A dominated nonlinear absorbtion
process for this wavelength is $K=5$ with $\beta^{(5)}=1.80\cdot
10^{-51} \text{cm}^{7}\text{W}^{-4}$
\cite{BergeSkupinNuterKasparianWolfBergePhysRep2007}. The nonlinear Kerr term in \eqref{nlsdimensionall}
dominates over the multiphoton absorbtion term provided the light
intensity $I<\left (\frac{2kn_{2}}{\beta^{(5)}n_0}\right )^{1/3}\equiv
I_{\rm MPA}\simeq3\cdot10^{13}\text{W}/\text{cm}^2$. The critical power
\eqref{Ncdef} in dimensional units $P_c=\frac{N_c
  \lambda_0^2}{8\pi^2n_2n_0}\simeq\text{2MW}.$

Assume that we propagate through the fused silica the collimated
Gaussian laser beam with the initial intensity distribution $I({\bf
  r},z=0)=I_{\rm ini}e^{-2r^2/w_0^2}$, where the initial beam waist
$w_0=0.5\text{cm}$. The beam power $I_{\rm ini}\pi w_0^2/2$ is just
above $P_c$. Then the initial beam intensity $I_{\rm ini}\simeq 6\cdot
10^6 \text{W}/\text{cm}^2.$ It means that the dynamic range of
intensities $I_{\rm MPA}/I_{\rm ini}\simeq 5\cdot 10^6$ of NLSE
applicability is quite large.  This estimate for $I_{\rm MPA}$ can be
considered as the upper upper limit of the allowed laser intensity.
This limit is valid for ultrashort optical pulse duration (tens of
fs). For longer pulses MPA eventually results in optical
damage. Typical experimental measurements of the optical damage
threshold give the threshold intensity $I_{\rm thresh}\sim 5\cdot
10^{11}\text{W}/\text{cm}^2$ for $8\,\text{ns}$ pulses and $I_{\rm
  thresh}\sim 1.5\cdot 10^{12}\text{W}/\text{cm}^2$ for $14\text{ps}$
pulses \cite{SmithDoApplOpt2008}. Even these lower estimates give more
than five orders of the dynamic range of NLSE applicability.  However,
for such short pulse durations, $t_0$, we generally might need to take
into account a group velocity dispersion (GVD).  Its contribution is
described by the addition of the term $-\frac{\beta_2}{2}
\frac{\partial^2}{\partial t^2}\tilde\psi$ into the left-hand side of
equation \eqref{nlsdimensionall}.  Here
$\beta_2=370\text{fs}^2/\text{cm}$ is the GVD coefficient for
$\lambda_0=790\text{nm}$ and $t$ is the retarded time $t\equiv T-z/c$,
where $T$ is the physical time and $c$ is the speed of light.  The
collapse distance, $\tilde z_c$, in dimensional units is given by
$\tilde z_c=\frac{4\pi n_0w_0^2}{\lambda_0}z_{c}\simeq 600\text{m}$,
where we set $z_c\simeq 1.047499$ as in the simulation shown in the
inset of Figure~\ref{fig:Lt}.  At this distance the linear absorbtion
of optical grade fused silica is still negligible.  The GVD distance
$\tilde z_{\rm GVD}\equiv 2t_0^2/\beta_2$ must exceed $\tilde z_c$ for
NLSE applicability, which gives $t_0\gtrsim 3\text{ps}$.

Another possible effects beyond NLSE include a stimulated Brillouin
scattering (can be neglected for the pulse duration $\lesssim
10\text{ns}$ \cite{AgrawalBookNonlinFiberOptics2012}) and a stimulated
Raman scattering (SRS).  The threshold of SRS for a long pulse in
fused silica was estimated from a gain exponent $gI_0l\simeq 16$, where the peak
intensity of the pulse, $I_0$, assumed to be constant along the
propagation distance $l$, and $g\simeq 10^{-11}\text{cm}/\text{W}$
is the Raman gain constant \cite{AgrawalBookNonlinFiberOptics2012}.
This estimate was obtained assuming that the spontaneous emission is
amplified by SRS (with the amplification factor $e^{gI_0l}=e^{16}$) up to the level of the laser pump intensity $I_0$ \cite{AgrawalBookNonlinFiberOptics2012}.
 In this paper we modify
this SRS threshold estimate to  account
for the variable pulse intensity along $z$ (the intensity evolves
according to \eqref{selfsimilar}).

The maximum of intensity at ${\bf r}=0$ evolves as
$I(z)\simeq\frac{L(z_{0})^2}{L(z)^2}I_{\rm ini}\simeq
\frac{z_c-z_{0}}{z_c-z}I_{\rm ini}$ for $z> z_0$, where $z_0$ is
defined above ($z_0\simeq 1.0007$ for the simulation of the inset of
Figure~\ref{fig:Lt}). Here we neglected the logarithmic contributions
to $L$, as well as we neglected a small contribution  to the total SRS amplification
 from the range $z<z_0$.  The SRS wave intensity $I_s$ is
amplified according to $\frac{d I_s(z)}{dz}=g I_s(z) I_{\rm
  ini}\frac{z_c-z_{0}}{z_c-z}$ which results in
$I_s(z_0+l)=I_s(z_0)\exp{\left ( gI_{\rm
    ini}(z_c-z_0)\ln{\frac{z_c-z_{0}}{z_c-z_0-l}} \right )}.  $ It
means that the collapse replaces the gain exponent $gI_{\rm ini}l$ (of
the constant intensity case $I_0=I_{\rm ini}$) by the modified gain
exponent $gI_{\rm ini}l\ln{\frac{z_c-z_{0}}{z_c-z_{0-}l}}$, where
$l\simeq z_c-z_0$. At SRS threshold that gain exponent has to be $\simeq 16$ as explained above. We now assume that the collapsing filament intensity
increases by 6 orders: $\frac{z_c-z_{0}}{z_c-z_0-l}=10^6.$ Then we
obtain the gain exponent $gI_{\rm ini}l\ln{\frac{z_c-z_{0}}{z_c-z_0-l}}\simeq 2\ll
16$,  i.e. we still operate well below the
SRS threshold and can neglect SRS.  This SRS threshold estimate is true for
relatively long pulses $\gtrsim 10\text{ps}$
\cite{AgrawalBookNonlinFiberOptics2012}.  For pulses of shorter
duration, SRS is additionally suppressed because the laser beam
and the SRS wave move with different group velocities.

We conclude that the optimal pulse duration for the experimental
verification of this paper is $3\text{ps}\lesssim t_0\lesssim
10\text{ns}$. Note that one can easily reduce the required media
length $\tilde z_c$ in several order of magnitude by prefocusing
of the pulse before it enters Kerr media.  However, the expense of such
prefocusing would be reduction of the dynamic range of NLSE
applicable intensities.

\section{Numerical simulations of NLSE}
\label{section:simulationalgorithm}

The results presented in this paper are obtained using an adaptive
mesh refinement (AMR) technique~\cite{Berger198964,SulemSulem1999},
complemented with the sixth-order Runge-Kutta time advancement
method. Some details of that type of technique are provided in
Ref. \cite{DyachenkoLushnikovVladimirovaKellerSegelArxiv2013}.  The
spatial derivatives are calculated using 8th order finite difference
scheme on the nonuniform grid.  Our spatial domain, $r\in
[0,r_{max}]$, is divided into several subdomains (subgrids) with
different spatial resolution. The spacing between computational points
is constant for each subgrid, and differs by a factor of two between
adjacent subgrids.  The rightmost subgrid, farthest from the collapse,
has the coarsest resolution; the spatial step decreases in the inward
direction. The grid structure adapts during the evolution of the
collapse to keep the solution well resolved.  When a refinement
condition is met, the leftmost subgrid is divided in two equal
subgrids with the interpolation of up to 10th order used to initialize
the data on the new subgrid.  The solution on all subgrids is evolved
with the same timestep, $\Delta t = C_{\rm CFL} h^2$, where $h$ is the
spatial step of the finest grid, and $C_{\rm CFL}$ is a
constant. Typically we choose $C_{\rm CFL}=0.05$, but we also tested
the convergence for smaller values of $C_{\rm CFL}$.

Finally, we comment on how we determine $L$ and $\beta$ from
numerical simulations.  At each $z$  we use the following two-step procedure.  First, we
determine $L(z)$ from the numerical solution $\psi(r,z)$ as
$L=\frac{1}{|\psi|}\left (1+2\frac{|\psi|_{rr}}{|\psi|^3}\right )^{-1/2}\Big |_{r=0}$,
an expression derived from the Taylor series expansion of $V_0(\beta,\rho)$
for $\rho \ll 1$ in \eqref{V0eq1}.  Second, we determine $\beta(z)$ from the nonlinear
condition $|\psi(0,z)|=\frac{1}{L(z)}V_0(\beta,0)$ using the
pre-computed values of $V_0(\beta,0)$ from the solution of
\eqref{V0eq1}.  We found that this procedure gives much better
accuracy in determining $L$ and $\beta$ than the
alternative procedures reviewed e.g. in
Ref. \cite{FibichPapanicolaouSIAMJApplMath1999}.

\section{Conclusion}
\label{sec:conclusion}

In conclusion, we found that the collapsing solution is described by
the approximate self-similar solution
$|\psi(r,z)|=\frac{1}{L(z)}V_0\left (\beta(z), \frac{r}{L(z)} \right
)$ for $0\le r/L(z)\lesssim 2/\beta^{1/2}(z)$ with $L(z)$ given by
\eqref{Ltot} and $\beta=-L^3 L_{zz}$, where $V_0(\beta,\rho)$ is the
ground state soliton solution of \eqref{V0eq1}.
The slow dependence  of $\beta$ on $z$ results in
adiabatically slow violation of self-similarity.
 For $r/L(z)\gg
2/\beta^{1/2}(z)$ the collapsing solution has the tail $\tilde V_0$
from \eqref{Vasymp}.  We found that the dependence $L(z)$ in
\eqref{Ltot} is in very good agreement with the direct numerical
simulations, as shown in Figure~\ref{fig:Lt}, starting from quite
moderate increase ($\sim 2-3$ times) of the amplitude of the initial
Gaussian beam. By the direct substitution of the values $L(z_0)$ and
$\beta(z_0)$ into \eqref{Ltot}, with $z_0$ defined in
Figure~\ref{fig:Lt}, one can see that expression~\eqref{Ltot} matches
the classical result \eqref{double} with accuracy $\sim 10-20\%$ only
for the unrealistically large amplitudes given by
\eqref{doubleexponent}. It suggests that the classical result
\eqref{double}, while being asymptotically correct, should be replaced
by much more accurate new formula \eqref{Ltot} for any currently
foreseeable physical systems.

\begin{acknowledgments}
This work was supported by the National Science Foundation grants DMS 0807131,  PHY 1004118 and PHY 1004110.
\end{acknowledgments}



\end{document}